\documentclass[twocolumn,secnumarabic,superscriptaddress,amssymb,nobibnotes,showpacs,aps,prd]{revtex4-1}

\usepackage{hyperref}
\usepackage{graphicx}
\usepackage{subfigure}

\begin{document}

\title{Medium effects on the quarkonia states above critical temperature}%

\author{Arpit Parmar}
\email{arpitspu@yahoo.co.in}
\affiliation{Department of Physics, Sardar Patel University, Vallabh Vidyanagar-388 120, India.}

\author{Bhavin Patel}
\email{azadpatel2003@yahoo.co.in}
\affiliation{Department of Physical Sciences, P D Patel Institute of Applied Sciences, Changa, India}

\author{P C Vinodkumar}
\email{pothodivinod@yahoo.com}
\affiliation{Department of Physics, Sardar Patel University, Vallabh Vidyanagar-388 120, India.}

\date{\today}%
\begin{abstract}
We present the quarkonia correlators for charmonium and bottomonium systems in the pseudoscalar, vector and scalar channels. For the description of quark-antiquark interaction we adopt the temperature dependant colour screening potential of the power law form. The spectroscopic parameters defined from the model are employed in the spectral functions to compute the quarkonia correlators. We find considerable medium modifications to the effective masses of the quarkonia as well as in the behaviour of the respective radial wave functions. These modifications are then reflected in the computed correlators. The general behaviour of correlators in the vector and scalar channel are in accordance with the latest lattice results while their behaviour in the pseudoscalar channels are found to be different.
\end{abstract}
\maketitle
\section{Introduction}
The behaviour of heavy quarkonium states (charmonium and bottomonium) in a hot and dense medium have attracted considerable experimental and theoretical interest. Heavy quarkonia play an important role in studying hot and dense strongly interacting matter. Study of the properties of heavy quarkonia above deconfinement temperature is of extreme interest for current Relativistic heavy ion collider (RHIC) experiments \cite{PhysRevD.76.094513,PPNP.65.209}. The topics of interest in these studies include survival probabilities as bound state at some temperature in quark gluon plasma and in medium transport properties of heavy and light quarks. Because of larger quark mass the heavy quarkonia can survive and remain in bound state above the deconfinement temperature $T_c$. The binding energy analysis has become important for such studies based on either potential models \cite{PhysRevD.73.074007} or by Lattice methods \cite{PhysRevD.69.094507,PhysRevD.86.014509,LAT.2005.153}.\\
The lattice results of quarkonia correlator studies predicts existence of $1S$ charmonium states upto $1.6T_c$ and of $1P$ states ($\chi_{c0}$ and $\chi_{c1}$) states upto $1.1T_c$. The present results on bottomonium study predicts existance of $1S$ states ($\eta_b$ and $\Upsilon$) upto temperatures $2.3T_c$ and for $1P$ states  ($\chi_b$) upto $1.15T_c$. Which are in contradiction with the earlier potential model calculations which predict dissolution of charmonium states below $1.1T_c$ \cite{ZPhysC.37.617,PhysRevD.64.094015}. With the large number of excited charmonia and bottomonia states known our earlier understanding based on potential models have been reviewed recently \cite{PhysRevD.73.074007} and systematic variations in the confinement strength has been introduced to explain the observed quarkonia spectra. Such a modification to the string tension of the confining part of the potential then be understood in terms of the medium effects. Based on such an attempt, here we extend our earlier works on the spectroscopy and decay properties of quarkonia using coulomb plus power law potential ($CPP_\nu$) model to a systematic study based on the quarkonia correlators for charm and bottom quark systems. To incorporate the medium effects on the binding energy of quarkonia we generalize the usual coulomb plus power law ($CPP_\nu$) form of the potential with a medium dependant exponential screening factor which reduces to the coulomb plus power law form in the absence of the medium effect. In section \ref{sec:correlators} we discuss the method to compute spectral functions and correlators. The spin average mass and the wave function dependant decay constants are derived in section \ref{sec:msa}. Finally, we discuss and summarize our results for quarkonia correlators against different choices of power exponent $\nu$ in section \ref{sec:results}.

\section{Euclidean correlators and spectral functions}\label{sec:correlators}

The temperature dependence of the meson correlators can provide information about the fate of quarkonia states above deconfinement. The imaginary time Euclidean correlation functions of meson currents $G(\tau,T)$ are reliably calculated on the lattice
\begin{equation}\label{eq:correlators}
G(\tau,T)=\int d\omega\sigma(\omega,T)K(\tau,\omega,T)
\end{equation}
Where, $\sigma(\omega,T)$ is the zero temperature spectral function and $K(\tau,\omega,T)$ is the kernel of integration and can be written as,
\begin{equation}
K(\tau,\omega,t)=\frac{cosh(\omega(\tau-\frac{1}{2T}))}{sinh(\frac{\omega}{2T})}
\end{equation}
In lattice QCD calculation the spectral function $\sigma(\omega)$ can be extracted out from the information of correlators using the maximum entropy method \cite{JPhysG.32.S515}. While, the potential models employ the spectral function to extract out the quarkonia correlators.
Following ref \cite{RevModPhys.65.1} the spectral function can be written as,

\begin{equation}\label{eq:spectralfunction}
\sigma(\omega)=\sum_i2M_iF_i^2\delta(\omega^2-M_i^2)+\frac{3}{8\pi^2}\omega^2\theta(\omega^2-s_0)f(\omega,s_0)
\end{equation}

The first term arises from the pole contributions from the bound states, and the second term is the perturbative continuum above some threshold $s_0$. Here, we consider the form of $f(\omega,s_0)$ motivated by leading order perturbative calculations with massive quarks as \cite{PhysRevD.73.074007},
\begin{equation}
f(\omega,s_0)=\left(a_H+b_H\frac{s_0^2}{\omega^2}\right)\sqrt{1-\frac{s_0^2}{\omega^2}}
\end{equation}
The calculated coefficients of $a_H$ and $b_H$ at leading order by \cite{PhysLettB.497.249} are shown in Table \ref{coefficients}. The value of threshold energy $s_0$ is somewhat arbitrary for single flavour heavy quark only. Here, $s_0$ is chosen such that no resonance above it is possible. The remaining parameters $M_i$ and wave function dependant decay constants $F_i$ are deduced from the potential model adopted for the present study.\\
To see the temperature effect on the spectral function and to compare with the lattice QCD results one usually computes the ratio of this correlators to the reconstructred  one $G(\tau,T)/G_{recon}(\tau,T)$, where $G_{recon}(\tau,T)$ is given by,

\begin{equation}
G_{recon}(\tau,T)=\int_0^\infty d\omega\sigma(\omega,T=0)K(\tau,\omega,T)
\end{equation}

Here, $G_{recon}(\tau,T)$ corresponds to the spectral function at the zero temperature. The temperature dependance in the $G_{recon}(\tau,T)$ comes only from the kernel of the integration.

\begin{table}[h]
\caption{The coefficients ($a_H,b_H$) in different mesonic channel \cite{PhysLettB.497.249}}\label{coefficients}
\begin{tabular}{lcc}
\hline
System&$a_H$&$b_H$\\
\hline
scalar&-1&1\\
pseudoscalar&1&0\\
vector&2&1\\
axial-vector&-2&3\\
\hline
\end{tabular}
\end{table}

\begin{figure}
\begin{center}
\includegraphics[width=1\linewidth,angle=0]{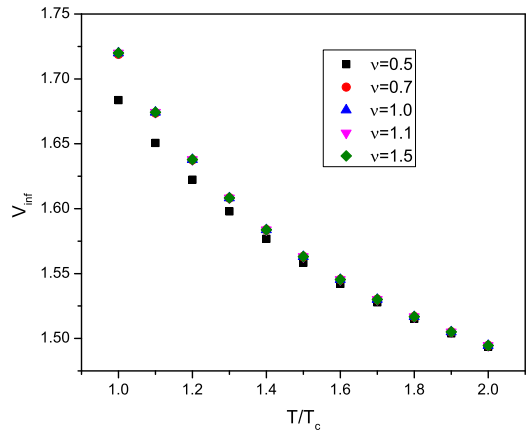}
\caption{$V(r=\infty,T)$ for different temperature at different potential exponent}\label{fig:vinf}
\end{center}
\end{figure}

\begin{figure*}
\begin{center}
\includegraphics[width=1\linewidth,angle=0]{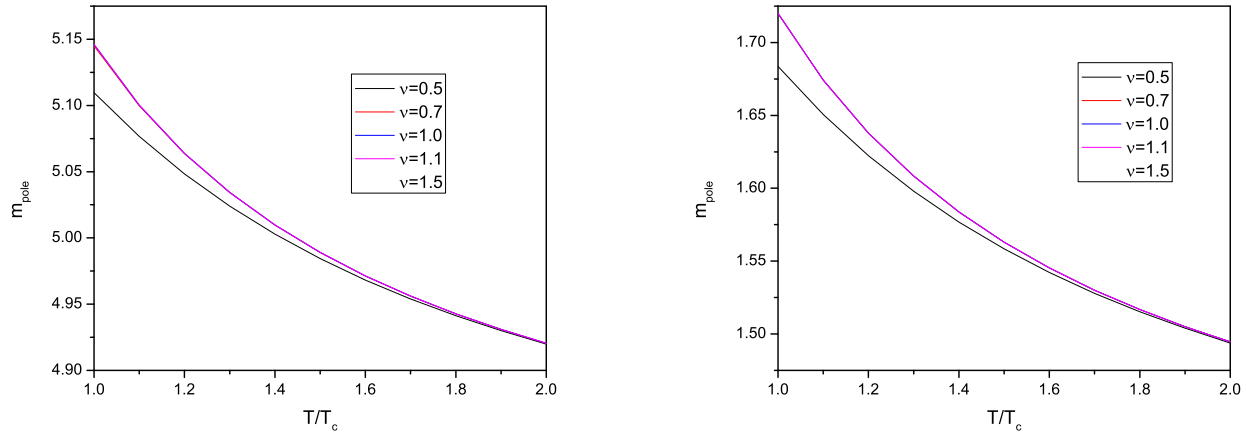}
\caption{The pole mass for bottom quark for screened coulomb potential at different potential exponent (a) for $b\bar b$ system (b) for $c\bar c$ system}\label{fig:mpolebb}
\end{center}
\end{figure*}

\begin{figure*}[!htbp]
\begin{center}
\includegraphics[width=1\linewidth,angle=0]{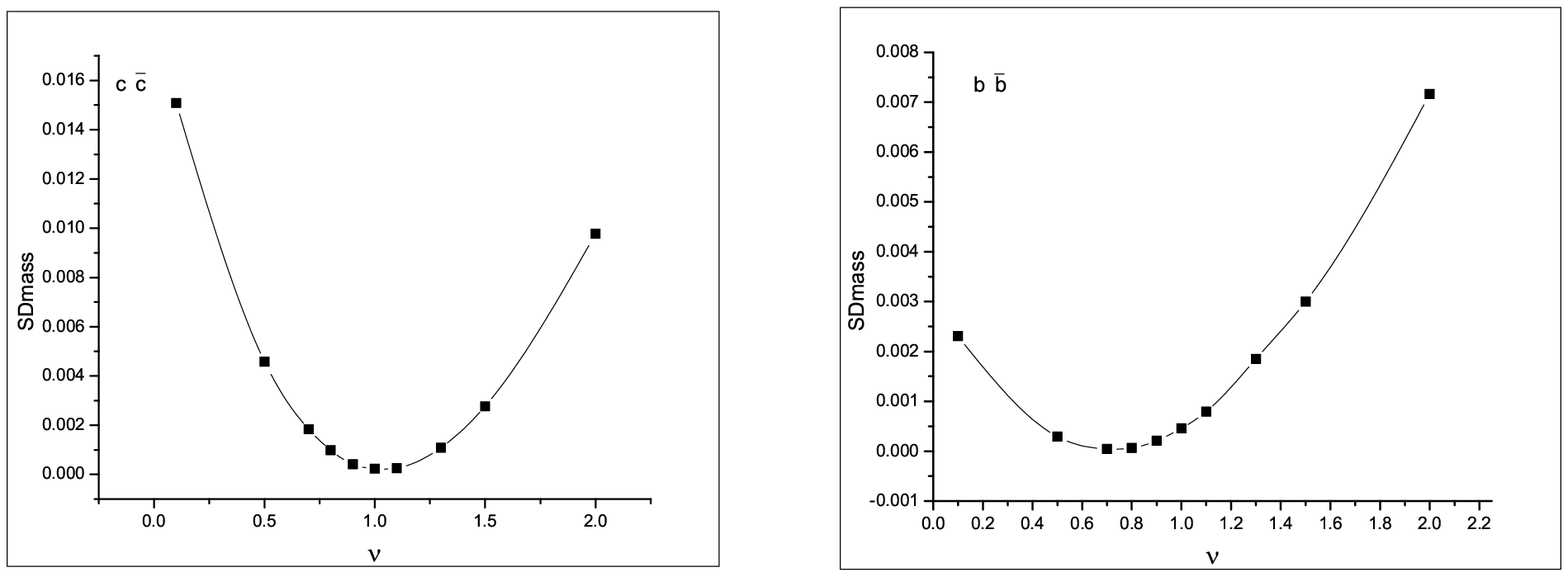}
\caption{rms deviation in zero temperature mass for quarkonia states taken from \cite{NuclPhysA.848.299} (a) for charmonium states (b) for bottomonium states}\label{fig:rmsmass}
\end{center}
\end{figure*}

\begin{figure*}
\begin{center}
\includegraphics[width=1\linewidth,angle=0]{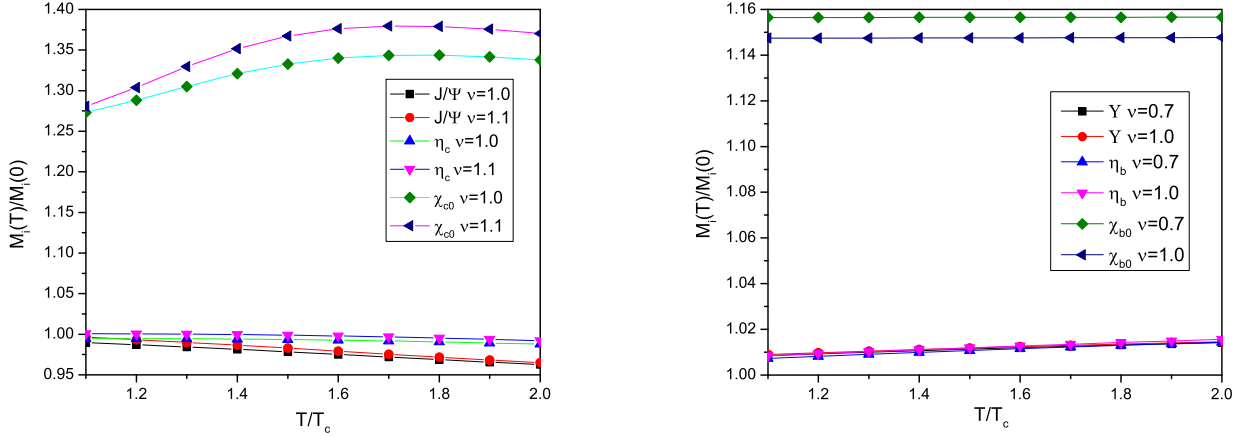}
\caption{Ratio of the mass at given temperature to the zero temperature mass for different quarkonium states at different potential exponent $\nu$ (a) for charmonium (b) for bottomonium.}\label{fig:ratiomass}
\end{center}
\end{figure*}

\begin{figure*}
\begin{center}
\includegraphics[width=1\linewidth,angle=0]{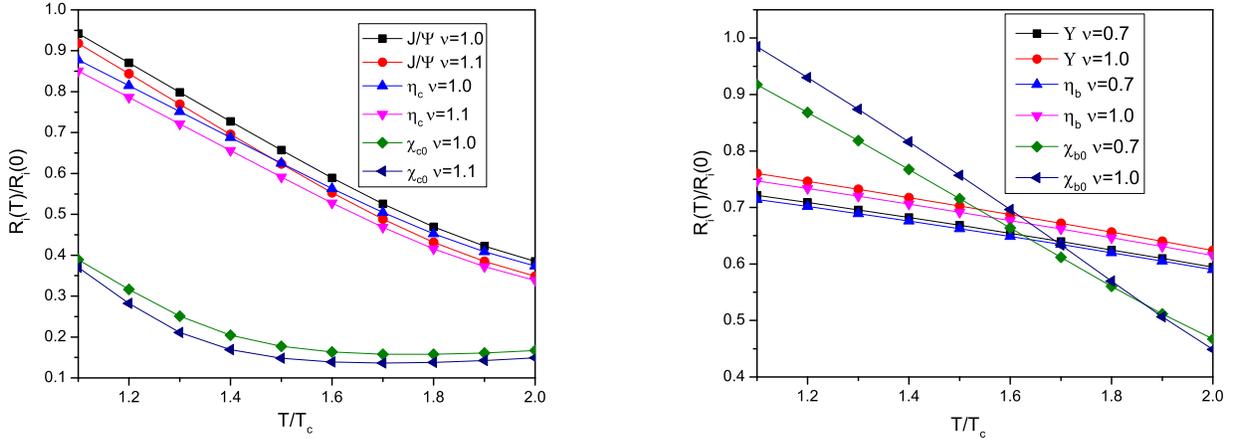}
\caption{Ratio of the wave function at zero separation  at given temperature $R_i(T)$ to the zero temperature wave function $R_i(0)$ for different quarkonium states at different potential exponent $\nu$ (a) for charmonium (b) for bottomonium.}\label{fig:ratiowf}
\end{center}
\end{figure*}

\begin{figure*}
\begin{center}
\includegraphics[width=1\linewidth,angle=0]{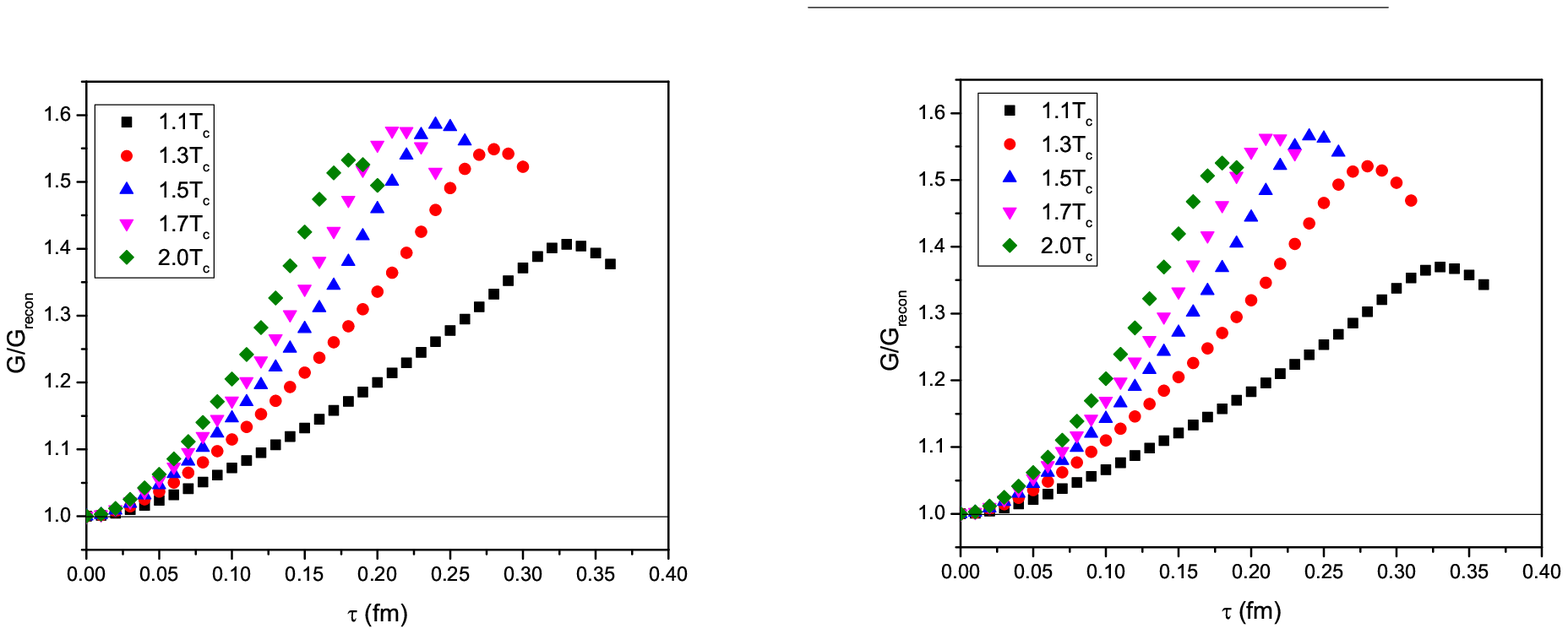}
\caption{$G/G_{recon}$ for pseudoscalar charmonium ($\eta_c$) state (a) for $\nu=1.0$ (b) for $\nu=1.1$}\label{fig:1sccp}
\end{center}
\end{figure*}

\begin{figure*}
\begin{center}
\includegraphics[width=1\linewidth,angle=0]{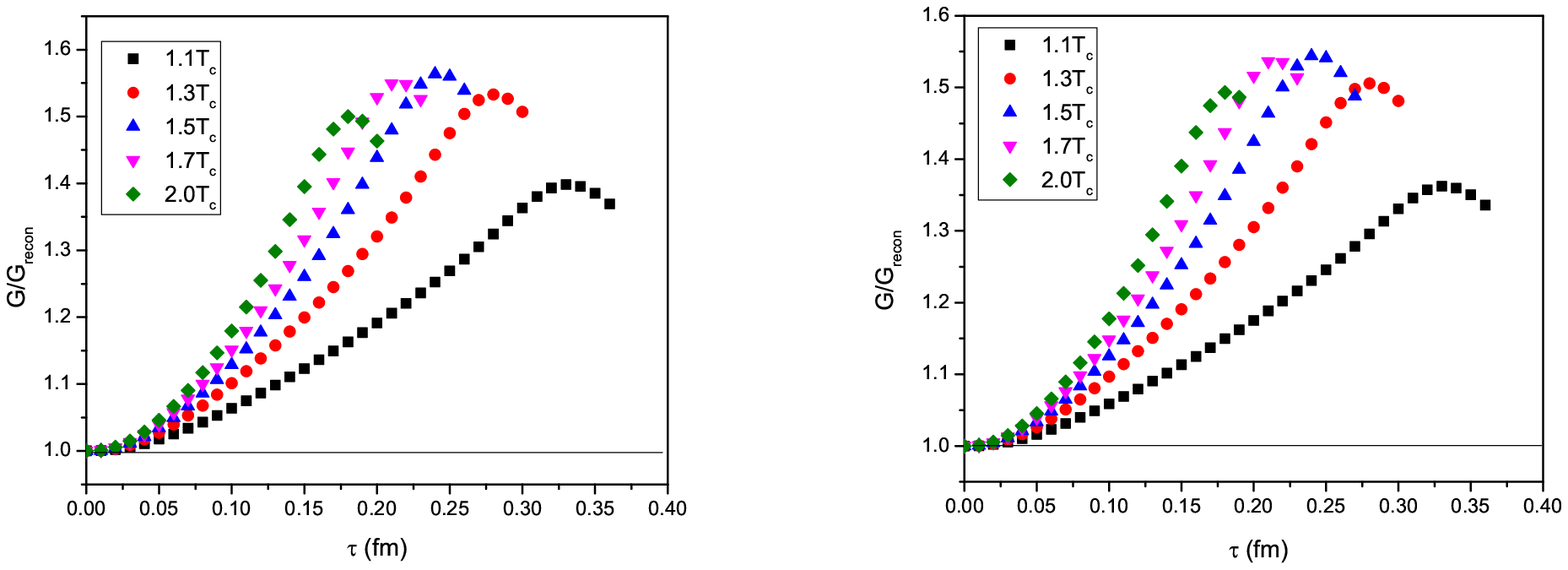}
\caption{$G/G_{recon}$ for vector charmonium ($J/\psi$) state (a) for $\nu=1.0$ (b) for $\nu=1.1$}\label{fig:1sccv}
\end{center}
\end{figure*}

\begin{figure*}
\begin{center}
\includegraphics[width=1\linewidth,angle=0]{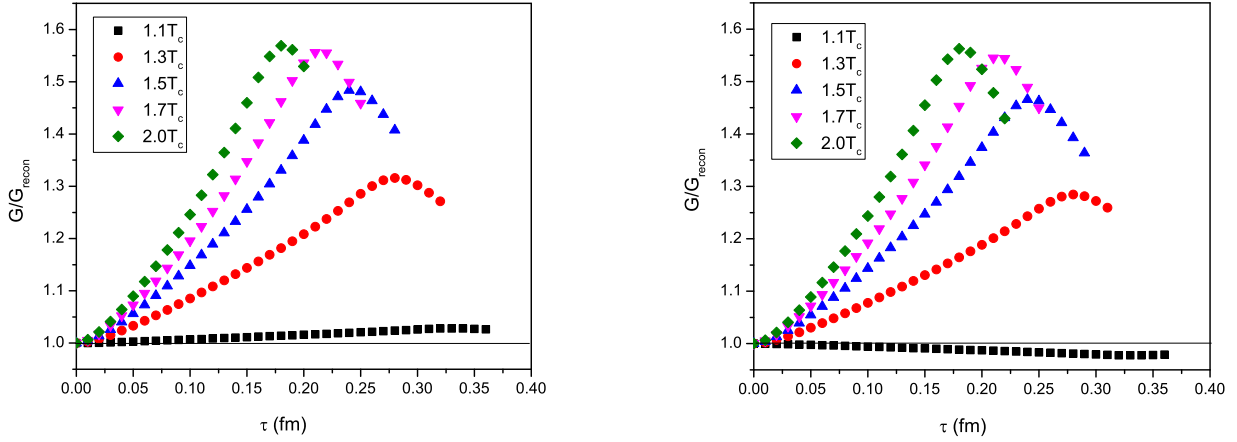}
\caption{$G/G_{recon}$ for scalar charmonium state ($\chi_{c0}$) (a) for $\nu=1.0$ (b) for $\nu=1.1$}\label{fig:1pccv}
\end{center}
\end{figure*}

\begin{figure*}
\begin{center}
\includegraphics[width=1\linewidth,angle=0]{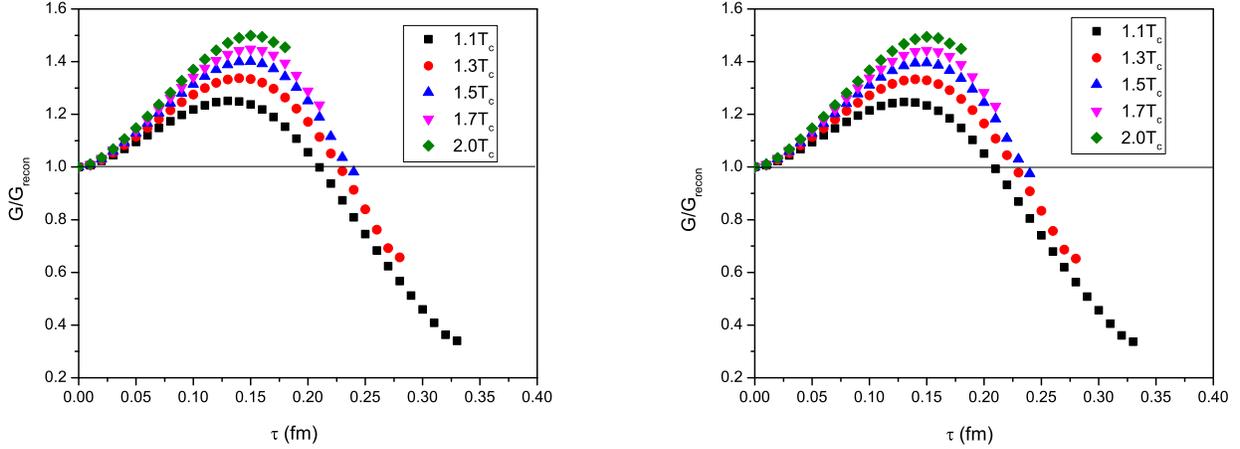}
\caption{$G/G_{recon}$ for pseudoscalar bottonium state ($\eta_b$) (a) for $\nu=0.7$ (b) for $\nu=1.0$}\label{fig:1sbbp}
\end{center}
\end{figure*}

\begin{figure*}
\begin{center}
\includegraphics[width=1\linewidth,angle=0]{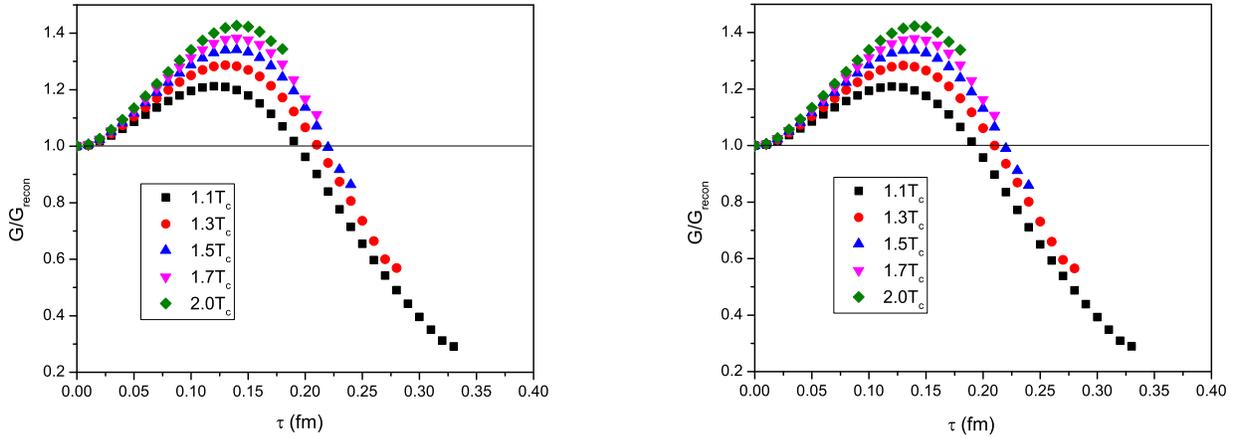}
\caption{$G/G_{recon}$ for vector bottonium state ($\Upsilon$) $\nu=0.7$ (a) for $\nu=0.7$ (b) for $\nu=1.0$}\label{fig:1sbbv}
\end{center}
\end{figure*}

\begin{figure*}
\begin{center}
\includegraphics[width=1\linewidth,angle=0]{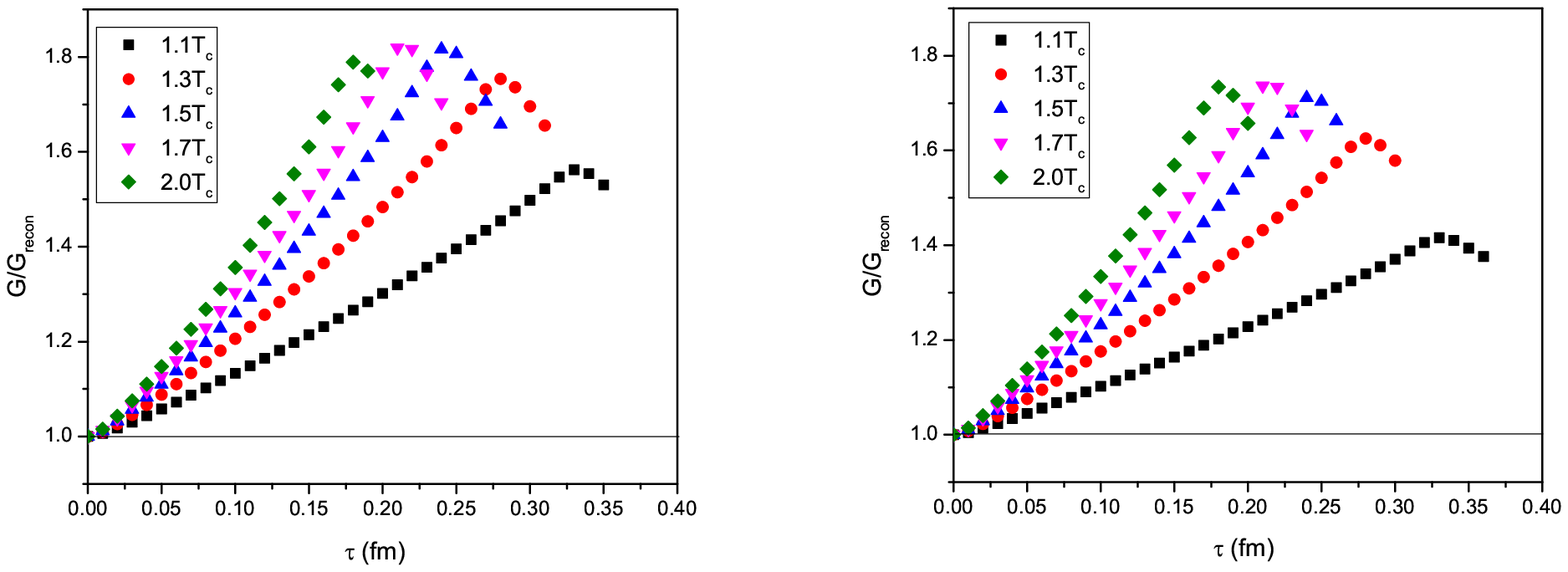}
\caption{$G/G_{recon}$ for scalar bottonium state ($\chi_{b0}$) (a) for $\nu=0.7$ (b) for $\nu=1.0$}\label{fig:1pbbv}
\end{center} \end{figure*}

\section{Extraction of spin average mass ($M_i$) and decay constants ($F_i$)}\label{sec:msa}
The spin average mass and decay constants appeared in the expression of the spectral function (Eqn. \ref{eq:spectralfunction}) are deduced using an appropriate model description of the hadronic state. Here, for the description of the quarkonia states we consider temperature dependant screened potential of the form,

\begin{equation}\label{eq:screenedpotential}
V(r,T)=-\frac{\alpha}{r}e^{-\mu(T)r}+\frac{\sigma}{A\mu(T)}(1-e^{-A\mu(T)r^\nu})
\end{equation}
Here, $\alpha$ and $\sigma$ are the coupling constant and string tension respectively. The exponent, $\nu$ is the variation that we have introduced here to account for the nonlinear dependance on the interquark distance of the potential. Different choices of the exponent, $\nu$ describes different strength of the interquark interaction. The screening mass parameter $\mu(T)$ is taken as, $\mu(T)=0.24+0.31(T/T_c-1)$ $GeV$ \cite{PhysRevD.73.074007} with critical temperature $T_c=0.270$ $GeV$. Here, we have chosen $A=1\ GeV^{\nu-1}$.\\
Here, the screening potential attains finite value at infinite separation i.e. $V(r=\infty,T)$. Hence, each quark has an additional thermal energy of $V_\infty/2$ in the bound state. The minimum energy above which the quark-antiquark pair can freely propagate is $2m+V_{\infty}$. Hence, the minimum energy threshold can be defined in this case as $s_0=2m+V_\infty$. The potential $V_\infty$ for different temperature is shown in Fig. \ref{fig:vinf} for different potential exponent. Similarly, the pole mass $m_i$ corresponds to $m_i=m+V_\infty/2$ against $T/T_c$ for different potential exponent $\nu$ are shown in Fig. \ref{fig:mpolebb} for bottomonium and charmonium states. $V_\infty$ found to be decreasing with increase in temperature above $T_c$. Similar observation is observed on lattice calculations while calculating quark and gluon propagators in coulomb gauge \cite{NuclPhysBProc.106.513}.
In the absence of the medium effect (at zero temperature) the potential in Eqn. \ref{eq:screenedpotential} reduces to the form
\begin{equation}
V(r)=-\frac{\alpha}{r}+\sigma r^\nu
\end{equation}
The parameters $\alpha$, $\sigma$ and the quark masses are fixed by fitting the zero temperature quarkonium spectrum as in \cite{PhysRevD.33.3338,NuclPhysA.848.299}. The parameters employed are given by $\alpha=0.471$, $\sigma=0.192$ $GeV^{\nu+1}$, $m_c=1.32$ $GeV$ and $m_b=4.746$ $GeV$ by fitting the zero temperature quarkonia spectrum.
The spin average masses and the wave functions are obtained by solving the schroedinger equation numerically using the screened potential of Eqn. \ref{eq:screenedpotential} as quark-antiquark interaction potential at a given temperature.

In the relativistic quark model, the decay constant can be expressed through the meson wave function $\Phi_{P/V}(p)$ in the momentum space \cite{PhysRevD.67.014027},
\begin{eqnarray}
f_{P/V} & = &\sqrt{\frac{12}{M_{P/V}}}\int\frac{d^3p}{(2\pi)^3} \sqrt{\left(\frac{E_Q(p) + m_Q}{2E_Q(p)}\right )\left ( \frac{E_{\bar Q}(p) + m_{\bar Q}}{2E_{\bar Q}(p)}\right )} \cr
&& \left \{1+ \frac{\lambda_{P/V}p^2}{[E_Q(p)+m_Q][E_{\bar Q}(p)+m_{\bar Q}]}\right \}\Phi_{P/V}(p)
\end{eqnarray}
with $\lambda_P=-1$ and $\lambda_V=-1/3$. In the nonrelativistic limit $\frac{p^2}{m^2}<<1.0$, this expression reduces to the well known relation between  $f_{P/V}$ and the ground state wave function at the origin $R_{P/V}(0)$ through the Van-Royen-Weisskopf formula \cite{Nu.Cim.50.}. Though most of the models predict the meson mass spectrum successfully, there exist wide range of predictions of their  decay constants. For example, the ratio $\frac{f_P}{f_V}$ was predicted to be $>1$ in most of the nonrelativistic cases, as $m_P<m_V$ and their wave function at the origin assumed to be  the same $R_P(0) \sim R_V(0)$ \cite{Z.Phys.76.107}. The ratio computed in the relativistic models \cite{PhysLettB.633.492} have predicted $\frac{f_P}{f_V}<1$, particularly in the $Q \bar Q$ sector, but $\frac{f_P}{f_V}>1$ in the heavy-light flavour sector. This disparity on the predictions of the decay constants play decisive role in the decay properties of these mesons. The value of the radial wave function ($R_{P}$) for $0^{-+}$ and ($R_{V}$) for $1^{--}$ states would be different due to their spin dependent hyperfine interaction. The spin hyperfine interaction of the heavy flavour mesons are small and this can cause a small shift in the value of the wave function at the origin. Though, many models neglect this difference between $(R_{P})$ and $(R_{V})$, we consider this correction by making an ansatz that the $R_{P/V}(0)$  are related to the value of the radial wave function at the origin, $R_{n}(0)$ according to the same way their masses are related. Thus, by considering
\begin{equation}\label{eq:Msa}
M_{nP/V}=M_{n,CW}\left[1+(SF)_{P/V} \frac{\langle V_{SS}\rangle_{n}}{M_{n,CW}}\right]
\end{equation}
and following the fact that any $c$-number, $a$, commutes with the Hamiltonian, \emph{i.e.} $aH\Psi=H(a\Psi)$, we express \cite{PhysRevC.78.055202},
\begin{equation}\label{Rcw1}
R_{nP/V}(0)=R_{n}(0)\left[1+(SF)_{P/V}\frac{(M_{nV}-M_{nP})}{M_{n,CW}} \right]
\end{equation}
Here $(SF)_P=-\frac{3}{4}$ and $(SF)_V=\frac{1}{4}$ are the spin factor corresponding to the pseudoscalar ($J=0$) spin coupling and vector ($J=1$) spin coupling respectively. $M_{n,CW}$ and $R_{n}(0)$ are spin average mass and the normalized spin independent wave function at the origin of the meson state respectively. It can easily be seen that this expression given by Eqn \ref{Rcw1} is consistent with the relation
\begin{equation}
R(0)=\frac{3 R_{V}(0)+R_{P}(0)}{4}
\end{equation}
given by \cite{PhysRevC.78.055202,PhysRevD.51.1125} for  $nS$ states. By incorporating first order QCD correction to the Van Royen-Weiskopff formula, the decay constant is computed as
\cite{ZPhysA.336.89,PhysRevD.52.181},
\begin{equation}
f^2_{P/V}(nS)=\frac{3 \left| R^{(\ell)}_{nP/V}(0)\right|^2} { \pi M_{nP/V}}
{\bar C^2}(\alpha_s) \label{eq:fpv}
 \end{equation}
where, the first order QCD correction, $\bar{C}(\alpha_{s}$) is expressed for the $Q \bar{Q}$ system as
\begin{equation}
{\bar C}(\alpha_s)=1-\frac{\alpha_s}{\pi}\delta^{V,P}\label{eq:c2}
\end{equation}
Here $\delta^{V} = \frac{8}{3}  $ \cite{ZPhysA.336.89,PhysRevD.52.181} and $\delta^P=2$ \cite{ZPhysA.336.89,PhysRevD.52.181,9803433}.

\begin{figure}
\begin{center}
\includegraphics[width=1\linewidth,angle=0]{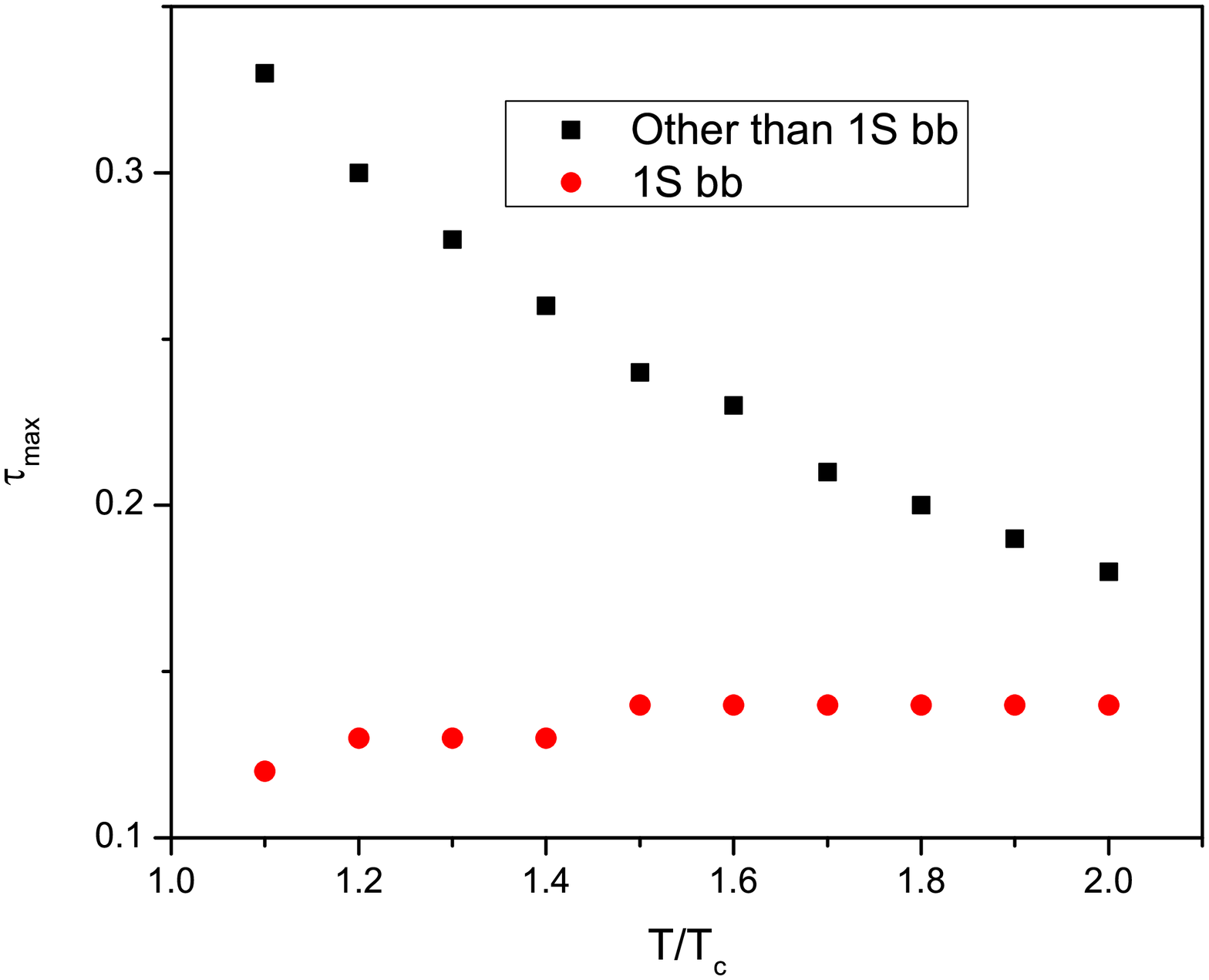}
\caption{$\tau_{max}$ vs. $T/T_c$ for all quarkonia states}\label{fig:taumax}
\end{center}
\end{figure}

\begin{table*}
\begin{center}
\caption{Spin averagem mass ($M_{sa}$) in $GeV$ for charmonium and bottomonium states for different choices of potential exponent $\nu$}\label{tab:msa}
\begin{tabular}{ccccccccccccccccc}
\hline
&\multicolumn{8}{c}{Charmonium state}&\multicolumn{8}{c}{Bottomonium state}\\
\cline{2-17}
&\multicolumn{4}{c}{1S}&\multicolumn{4}{c}{1P}&\multicolumn{4}{c}{1S}&\multicolumn{4}{c}{1P}\\
\cline{2-17}
$T/T_c$&$\nu=$0.5&1.0&1.1&1.5&0.5&1.0&1.1&1.5&0.5&0.7&1.0&1.5&0.5&0.7&1.0&1.5\\
\hline
1.1&2.930&3.040&3.060&3.129&3.063&3.274&3.306&3.370&9.515&9.519&9.525&9.535&9.787&9.831&9.891&9.976\\
1.2&2.929&3.034&3.053&3.115&3.053&3.242&3.268&3.305&9.523&9.527&9.533&9.542&9.787&9.829&9.887&9.967\\
1.3&2.928&3.028&3.045&3.100&3.042&3.211&3.229&3.254&9.530&9.534&9.540&9.549&9.786&9.827&9.883&9.958\\
1.4&2.926&3.021&3.036&3.085&3.033&3.181&3.195&3.212&9.537&9.541&9.547&9.555&9.784&9.824&9.878&9.947\\
1.5&2.923&3.013&3.027&3.070&3.024&3.154&3.165&3.179&9.544&9.548&9.554&9.562&9.782&9.821&9.872&9.935\\
1.6&2.921&3.005&3.018&3.054&3.018&3.132&3.141&3.153&9.550&9.554&9.560&9.568&9.780&9.817&9.865&9.922\\
1.7&2.918&2.997&3.008&3.039&3.013&3.113&3.121&3.131&9.556&9.560&9.566&9.573&9.776&9.813&9.858&9.907\\
1.8&2.915&2.989&2.999&3.025&3.009&3.099&3.105&3.114&9.562&9.566&9.572&9.579&9.773&9.808&9.850&9.892\\
1.9&2.912&2.980&2.989&3.012&3.007&3.087&3.093&3.101&9.568&9.572&9.577&9.584&9.769&9.803&9.842&9.876\\
2.0&2.910&2.973&2.981&3.000&3.007&3.079&3.084&3.091&9.573&9.577&9.582&9.589&9.766&9.797&9.834&9.862\\
\hline
\end{tabular}
\end{center}
\end{table*}

\begin{table*}
\begin{center}
\caption{Masses of charmonium states in $GeV$ for different choices of potential exponent $\nu$}\label{tab:mcc}
\begin{tabular}{ccccccccccccc}
\hline
&\multicolumn{4}{c}{$\eta_c$}&\multicolumn{4}{c}{$J/\psi$}&\multicolumn{4}{c}{$\chi_{c0}$}\\
\cline{2-13}
$T/T_c$&$\nu=$0.5&1.0&1.1&1.5&0.5&1.0&1.1&1.5&0.5&1.0&1.1&1.5\\
\hline
1.1&2.886&2.960&2.973&3.020&2.945&3.067&3.089&3.166&4.312&4.412&4.469&5.482\\
1.2&2.888&2.960&2.973&3.017&2.943&3.059&3.079&3.148&4.323&4.464&4.549&5.930\\
1.3&2.889&2.959&2.972&3.014&2.940&3.051&3.069&3.129&4.332&4.523&4.639&6.197\\
1.4&2.890&2.958&2.970&3.010&2.938&3.041&3.058&3.110&4.337&4.578&4.716&6.322\\
1.5&2.890&2.957&2.968&3.006&2.934&3.032&3.047&3.091&4.337&4.619&4.770&6.350\\
1.6&2.890&2.955&2.965&3.000&2.931&3.022&3.035&3.072&4.334&4.645&4.801&6.318\\
1.7&2.890&2.952&2.962&2.993&2.927&3.012&3.024&3.054&4.327&4.656&4.814&6.249\\
1.8&2.889&2.948&2.957&2.985&2.924&3.002&3.012&3.038&4.319&4.657&4.812&6.160\\
1.9&2.887&2.944&2.952&2.977&2.921&2.993&3.002&3.023&4.309&4.650&4.800&6.059\\
2.0&2.886&2.940&2.947&2.969&2.918&2.984&2.992&3.010&4.297&4.637&4.781&5.955\\
\hline
\end{tabular}
\end{center}
\end{table*}

\begin{table*}
\begin{center}
\caption{Masses of bottomonium states in $GeV$ for different choices of potential exponent $\nu$}\label{tab:mbb}
\begin{tabular}{ccccccccccccc}
\hline
&\multicolumn{4}{c}{$\eta_b$}&\multicolumn{4}{c}{$\Upsilon$}&\multicolumn{4}{c}{$\chi_{b0}$}\\
\cline{2-13}
$T/T_c$&$\nu=$0.5&0.7&1.0&1.5&0.5&0.7&1.0&1.5&0.5&0.7&1.0&1.5\\
\hline
1.1&9.464&9.464&9.464&9.465&9.532&9.538&9.546&9.558&11.359&11.359&11.360&11.361\\
1.2&9.472&9.472&9.472&9.474&9.540&9.545&9.553&9.565&11.359&11.359&11.360&11.361\\
1.3&9.481&9.480&9.481&9.482&9.547&9.552&9.560&9.571&11.359&11.359&11.360&11.361\\
1.4&9.489&9.488&9.489&9.490&9.554&9.559&9.567&9.577&11.359&11.360&11.360&11.362\\
1.5&9.496&9.496&9.497&9.498&9.560&9.565&9.573&9.583&11.360&11.360&11.360&11.363\\
1.6&9.504&9.504&9.504&9.505&9.566&9.571&9.579&9.589&11.360&11.360&11.361&11.363\\
1.7&9.511&9.511&9.511&9.512&9.572&9.577&9.584&9.594&11.360&11.360&11.361&11.365\\
1.8&9.517&9.518&9.518&9.519&9.577&9.582&9.589&9.599&11.360&11.360&11.361&11.366\\
1.9&9.524&9.524&9.525&9.526&9.582&9.587&9.594&9.603&11.360&11.361&11.362&11.367\\
2.0&9.530&9.531&9.531&9.532&9.587&9.592&9.599&9.608&11.360&11.361&11.362&11.369\\
\hline
\end{tabular}
\end{center}
\end{table*}

\begin{table*}
\begin{center}
\caption{$\ell^{th}$ derivative of radial wave function at zero separation for charmonium states in $GeV^{\ell+3/2}$ for different choices of potential exponent $\nu$}\label{tab:wfcc}
\begin{tabular}{ccccccccccccc}
\hline
&\multicolumn{4}{c}{$\eta_c$}&\multicolumn{4}{c}{$J/\psi$}&\multicolumn{4}{c}{$\chi_{c0}$}\\
\cline{2-13}
$T/T_c$&$\nu=$0.5&1.0&1.1&1.5&0.5&1.0&1.1&1.5&0.5&1.0&1.1&1.5\\
\hline
1.1&0.239&0.423&0.456&0.565&0.249&0.454&0.492&0.620&0.051&0.102&0.106&0.047\\
1.2&0.224&0.392&0.421&0.509&0.232&0.419&0.452&0.554&0.046&0.083&0.080&0.033\\
1.3&0.208&0.362&0.387&0.453&0.216&0.385&0.412&0.488&0.042&0.066&0.060&0.029\\
1.4&0.193&0.331&0.352&0.396&0.200&0.350&0.373&0.423&0.039&0.054&0.048&0.028\\
1.5&0.179&0.301&0.317&0.341&0.184&0.317&0.334&0.361&0.037&0.047&0.042&0.028\\
1.6&0.165&0.271&0.283&0.290&0.170&0.284&0.296&0.304&0.037&0.043&0.040&0.029\\
1.7&0.154&0.243&0.251&0.247&0.158&0.253&0.262&0.257&0.037&0.042&0.039&0.030\\
1.8&0.144&0.218&0.223&0.212&0.147&0.226&0.231&0.220&0.038&0.041&0.039&0.032\\
1.9&0.136&0.197&0.200&0.187&0.139&0.203&0.206&0.193&0.040&0.042&0.041&0.035\\
2.0&0.130&0.180&0.181&0.169&0.133&0.185&0.187&0.174&0.042&0.044&0.042&0.038\\
\hline
\end{tabular}
\end{center}
\end{table*}

\begin{table*}
\begin{center}
\caption{$\ell^{th}$ derivative of radial wave function at zero separation for bottomonium states in $GeV^{\ell+3/2}$ for different choices of potential exponent $\nu$}\label{tab:wfbb}
\begin{tabular}{ccccccccccccccccc}
\hline
&\multicolumn{4}{c}{$\eta_b$}&\multicolumn{4}{c}{$\Upsilon$}&\multicolumn{4}{c}{$\chi_{b0}$}\\
\cline{2-13}
$T/T_c$&$\nu=$0.5&0.7&1.0&1.5&0.5&0.7&1.0&1.5&0.5&0.7&1.0&1.5\\
\hline
1.1&6.810&7.385&8.147&9.182&6.909&7.502&8.289&9.363&0.527&0.671&0.874&1.168\\
1.2&6.695&7.258&8.001&9.005&6.791&7.370&8.138&9.179&0.498&0.635&0.826&1.089\\
1.3&6.575&7.125&7.852&8.826&6.668&7.234&7.983&8.993&0.468&0.598&0.776&1.006\\
1.4&6.451&6.990&7.698&8.642&6.540&7.095&7.825&8.802&0.437&0.561&0.725&0.917\\
1.5&6.322&6.851&7.542&8.455&6.407&6.951&7.663&8.608&0.407&0.524&0.672&0.822\\
1.6&6.189&6.707&7.381&8.266&6.271&6.803&7.497&8.412&0.376&0.485&0.618&0.715\\
1.7&6.054&6.560&7.217&8.072&6.132&6.652&7.328&8.211&0.347&0.447&0.563&0.597\\
1.8&5.915&6.409&7.051&7.876&5.989&6.497&7.157&8.008&0.319&0.410&0.506&0.483\\
1.9&5.771&6.257&6.881&7.676&5.841&6.340&6.981&7.802&0.294&0.374&0.450&0.392\\
2.0&5.626&6.100&6.707&7.475&5.693&6.179&6.803&7.594&0.273&0.342&0.398&0.331\\
\hline
\end{tabular}
\end{center}
\end{table*}

\section{Temperature and the exponent ($\nu$) dependance on the correlators}
The temperature dependant energy eigen values ($E_i$)and the wave functions are obtained by solving the schrodinger equation numerically \cite{IJMPC.10.607}. The spin average mass ($M_{sa}$) is given by $M_{sa}=2m_i+E_i$. Once $M_{sa}$ is obtained, then the mass of the quarkonia state is computed using Eqn. \ref{eq:Msa}. For computing the mass difference between different spin degenerate mesonic states, we consider the spin dependent part of the usual one
gluon exchange potential (OGEP) given by
 \cite{PhysRevD.72.054026,PhysRevD.74.014012,PPNP.61.455,RevModPhys.80.1161,PhysRevD.51.3613}.
Accordingly, the spin-dependent part, $V_{SD}(r)$ contains three types of interaction terms, such as the spin-spin, the spin-orbit and the tensor part as
\begin{eqnarray}\label{spin}
                V_{SD}(r) &=&  V_{SS}(r)\left[S(S+1)-\frac{3}{2}\right]+
                          V_{LS}(r)\left(\vec{L}\cdot\vec{S}\right)+\cr
               && V_{T}(r) \left[S
(S+1)-\frac{3(\vec{S}\cdot\vec{r})(\vec{S}\cdot\vec{r})}{r^2}\right]
              \end{eqnarray}
The spin-orbit term containing $V_{LS}(r)$ and the tensor term containing $V_{T}(r)$ describe the fine structure of the meson states, while the spin-spin term containing $V_{SS}(r)$ proportional to
$2(\vec{s_{q}}\cdot\vec{s_{\bar q}})=S(S+1)-\frac{3}{2} $ gives the spin singlet-triplet hyperfine splitting. The coefficient of these spin-dependent terms of Eqn.\ref{spin} can be written in terms of the vector ($V_V$) or coulomb part and scalar ($V_S$) or confinement part of the static potential, $V(r)$ described in Eqn. \ref{eq:screenedpotential} as \cite{PPNP.61.455}
\begin{equation} \label{LS}
V_{LS}(r)=\frac{1}{2\ m_{1} m_{2}\
r}\left(3\frac{dV_{V}}{dr}-\frac{dV_{S}}{dr}\right)
\end{equation}
\begin{equation} \label{T}
V_{T}(r)=\frac{1}{6\ m_{1} m_{2}
}\left(3\frac{d^{2}V_{V}}{dr^{2}}-\frac{1}{r}\frac{dV_{V}}{dr}\right)
\end{equation}
\begin{equation} \label{SS}
V_{SS}(r)=\frac{16 \ \pi
\alpha_{s}}{9\ m_{1} m_{2} } \delta^{(3)}(\vec{r})
\end{equation}

The spin average masses ($M_{sa}$) for both charmonium and bottomonium states are tabulated in Table \ref{tab:msa}. The quarkonia masses for both charmonium and bottomonium states are shown in Table \ref{tab:mcc} and \ref{tab:mbb} respectively. Similarly the $\ell^{th}$ derivative of the radial wave function at zero separation for charmonium and bottomonium states are shown in Table \ref{tab:wfcc} and \ref{tab:wfbb} respectively. The choices of the potential exponent $\nu=1.1$ for charmonium state and $\nu=0.7$ for bottomonium state correspond to the minimum of the standard deviation in the mass ($SD_{min}$) in zero temperature potential as observed in our previous study \cite{NuclPhysA.848.299} as shown in Fig. \ref{fig:rmsmass}. The temperature dependant masses of the quarkonia states in terms of their zero temperature masses ($M_i(T)/M_i(0)$) are computed for $T>T_c$ and for different choices of the potential exponent, $\nu$ are shown in Fig. \ref{fig:ratiomass}, similarly the ratio for the wavefunction is shown in Fig. \ref{fig:ratiowf}. The ratio in mass for the P-state is found to be larger compare to the S-state quarkonia states. The ratio in the wave function decreases with increasing tempearature for all quarkonium states. Finally, the decay constants are computed with help of Eqn. \ref{eq:fpv}.\\
Our computed Spin average masses ($M_{sa}$) and the decay constants $(f_{p/v})$ corresponding to the choises of $\nu=1.1$ in the charmonia and $\nu=0.7$ in the case of bottomonia are then employed to predict the spectral functions given in Eqn. \ref{eq:spectralfunction}. We have also computed spectral function for the choices $\nu=1.0$ in both the cases for the purpose of comparison with cornel like interactions considered by others \cite{PhysRevD.73.074007,PhysRevD.82.054008}. Further the euclidean correlators are computed using Eqn. \ref{eq:correlators}. The present ratio $G/G_{recon}$ are drawn in Figs. \ref{fig:1sccp}, \ref{fig:1sccv}, \ref{fig:1pccv}, \ref{fig:1sbbp}, \ref{fig:1sbbv} and \ref{fig:1pbbv} for the euclidean time $\tau$ in between $\tau=0$ to $\tau=1/T$. The plots of these figures correspond to different choice of $T/T_c\leq2.0$.

\section{Results and Discussion}\label{sec:results}
As from Fig. \ref{fig:ratiomass} $M_i(T)/M_i(0)$ for $J/\psi$ and $\eta_c$ case found to decrease continuously with $T/T_c$ for all the range of $T/T_c$. The behaviour is same for both the choices of $\nu=1.0$ and $\nu=1.1$ except the magnitude of $M_i(T)/M_i(0)$ increases slightly with $\nu$. While $M_i(T)/M_i(0)$ for $\chi_{c0}$ found to increase upto$T/T_c=1.7$ then, start slightly decreasing. However, in the case of $b\bar b$  system $M_i(T)/M_i(0)$ for all $\eta_b$, $\Upsilon$ and $\chi_{b0}$ states found to increases with $T/T_c$ for all $T$ in the range of $T/T_c\leq2.0$. So, the medium effect in the $\eta_c$ and $J/\psi$ are opposite to that of $\eta_b$ and $\Upsilon$ states.\\
From, Fig. \ref{fig:ratiowf} $R_i(T)/R_i(0)$ for $\eta_c$ and $J/\psi$ states found to decrease monotonically with increase in $T/T_c$, while $R_i^{(\ell)}(T)/R_i^{(\ell)}(0)$ for $\chi_{c0}$ state found to decrease upto $T/T_c=1.7$ then it is increasing slightly. $R_i(T)/R_i(0)$ for $\eta_b$ and $\Upsilon$ states found to decrease monotonically for all the range of $T/T_c$. $R_i^{(\ell)}(T)/R_i^{(\ell)}(0)$ for $\chi_{b0}$ state also found to decrease much faster than $\eta_b$ and $\Upsilon$ states. These observations are same for both $\nu=1.0$ and $\nu=0.7$ except their magnitudes are higher.\\
The correlators at the different $T/T_c$ obtained are found to have a similar behaviour with increase in $\tau$. In the case of charmonia state, correlators attain maxima at different values of $\tau$ and it shilfts towards lower $\tau$ as $T/T_c$ increases. While in the case of $\eta_b$ and $\Upsilon$ bottomonia the maxima of the correlators are found to shift slightly towards higher $\tau$. Whereas, the behaviour in the case of $\chi_{b0}$ belongs to that of charmonia states. To make this boservation more clear we plot the values of $\tau$ at the maximum correlator ($tau_{max}$) against $T/T_c$ in Fig. \ref{fig:taumax} for both the case of $b\bar b$ and $c\bar c$. It is very striking to see that values of $\tau$ at which $G/G_{recon}$ is maximum ($\tau_{max}$) is same for $\eta_c$, $J/\psi$, $\chi_{c0}$ and $\chi_{b0}$ states with a chosen $T/T_c$, and $\tau_{max}$ decreases with $T/T_c$. This behaviour is shown graphically in Fig. \ref{fig:taumax} Here we also shown the behaviour of $\tau_{max}$ vs $T/T_c$ in the case of $\eta_b$ and $\Upsilon$ system. It seems that both the curves leading towards a common saturated value beyond $T/T_c>2.0$. Our results do not show any major deviation with the potential exponent $\nu$. However, from Fig. \ref{fig:1pccv} (b), we find $G/G_{recon}$at $T/T_c=1.1$ of $\chi_{c0}$ decreases from 1.0 as $\tau$ increases while similar case in Fig. \ref{fig:1pccv} (a) for $\nu=1.0$ seems to increase from 1.0 as $\tau$ increases. From Fig. \ref{fig:1sbbv} and Fig. \ref{fig:1sbbp} we also notice that $G/G_{recon}$ in $\eta_b$ and $\Upsilon$ decreases below 1.0 relatively at lower $\tau$ values compared to all other states studied here. The $\tau$ dependance of the correlators are found to be sensitive to the choices of the continuum threshold $s_0$.\\
As seen from Fig. \ref{fig:1sccp} our obtained $G/G_{recon}$ shows agreement with the early potential model \cite{PhysRevD.73.074007} but differs from the lattice results \cite{PhysRevD.86.014509} where $G/G_{recon}$ remains to unity upto $\tau\approx0.05\ fm$ then it starts gradually falling. Our results shows same trend to QCD sum rule results \cite{PhysRevD.82.054008} at $T=1.1T_c$.\\
Fig. \ref{fig:1sccv} represents $G/G_{recon}$ for vector charmonium states ($J/\psi$). For smaller values of $\tau$ our results agrees with the lattice \cite{PhysRevD.86.014509} and also with the QCD sum rules \cite{PhysRevD.82.054008}.  Though the present $G/G_{recon}$ in the case of $\eta_c$ increases from 1.0, the results based on fine isotropic lattice study shows decreases from 1.0 for all temepratures. However the general behaviour of all other states are in accordance with the lattice results \cite{PhysRevD.86.014509,PhysRevD.69.094507}. As, the $\eta_c$ and $J/\psi$ corresponds to same 1S state, one do not expect their correlators to behave differently.\\
Fig. \ref{fig:1pccv} represents $G/G_{recon}$ for scalar channel of 1P charmonium state ($\chi_{c0}$). Here, also our results agrees with lattice \cite{PhysRevD.86.014509} and QCD sum rules \cite{PhysRevD.82.054008}.\\
Similarly, Figs. \ref{fig:1sbbp}, \ref{fig:1sbbv} and \ref{fig:1pbbv} represent our results for 1S pseudoscalar ($\eta_b$), 1S vector ($\Upsilon$) and 1P scalar ($\chi_{b0}$) bottomonium states respectively. Our results agrees with earlier potential model \cite{PhysRevD.73.074007} but contradicts with lattice results \cite{LAT.2005.153}. The lattice results for $\eta_b$ case shows unity behaviour for temperature upto $2.30T_c$. The weaker dependance of the correlators on temperature was assumed as the melting of the states, but melting of state cannot be exctracted from the correlators \cite{EurPhysJC.71.1534}. Similar to the charmonium case, our results for $\eta_b$ and $\Upsilon$ shows the same behaviour as expected (because they belong to same 1S state of bottomonium system).\\
Variation in $G/G_{recon}$ from unity represents dissolution of the quarkonium states into medium at higher temperature above $T_c$. Though, the lattice results for bottomonium states need to reexamined as the cutoff energy remains below the energy relevant to bottomonium systems \cite{PhysRevD.73.074007}.\\
$\tau$ for maximum occurance of $G/G_{recon}$, $\tau_{max}$ represents the temperature for the maximum correlation of the quarkonia states. In our study, $\tau_{max}$ found to be exponentially decreasing with the temperature in the case of quarkonia states (except for $b\bar b(1S)$ case).\\
The values of $G/G_{recon}$ below 1 corresponds to value of $G$ to be lower than $G_{recon}$. This means the survival probability of the state to be lower than in the zero mode case or dissociation of the quarkonium state. Considering the time for $G/G_{recon}=1$ related with the temperature T as, $\tau=1/T$ we predict the dissociation temperature of $\eta_c$, $J/\psi$ and $\chi_{c0}$ to be around $1.2T_c$ while $\eta_b$ and $\Upsilon$ states can survive upto $3.0T_c$.

\section*{Acknowledgement}
Part of this work is carried out under the UGC grant with ref no. F.40-457/2011(SR). Arpit Parmar thanks UGC, India for the financial support under RFSMS scheme for the present work.

\bibliographystyle{apsrev4-1.bst}
\bibliography{<your-bib-database>}

\end{document}